\documentclass[journal=jpcb,manuscript=article,layout=twocolumn]{achemso}
\usepackage{siunitx}
\SectionNumbersOn
\usepackage[version=3]{mhchem} 
\usepackage{setspace}
\usepackage[symbol]{footmisc}

\author{Yuvraj Singh}
\affiliation[NYUChem]{Department of Chemistry, New York University}
\author{Glen M. Hocky}
\email{hockyg@nyu.edu}
\affiliation[NYUChem]{Department of Chemistry, New York University}
\alsoaffiliation[SCCPC]{Simons Center for Computational Physical Chemistry, New York University}

\usepackage[capitalize]{cleveref}
\newcommand{\fmin}{ {F_\mathrm{min}} }
\newcommand{\fmax}{ {F_\mathrm{max}} }
\renewcommand{\vec}{\mathbf}
\usepackage{natmove}
\usepackage{xcolor}

\title[FISST+RX]{{Improved prediction of molecular response to pulling by combining force tempering with replica exchange methods}}

\begin{document}

\begin{tocentry}
\includegraphics[]{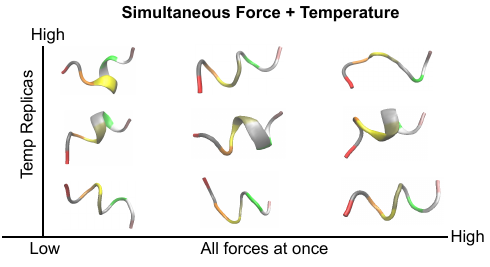}
\end{tocentry}

\begin{abstract}
Small mechanical forces play important functional roles in many crucial cellular processes, including in the dynamical behavior of the cytoskeleton and in the regulation of osmotic pressure through membrane-bound proteins. 
Molecular simulations offer the promise of being able to design the behavior of proteins that sense and respond to these forces.
However, it is difficult to predict and identify the effect of the relevant piconewton (pN) scale forces due to their small magnitude.
Previously, we introduced the Infinite Switch Simulated Tempering in Force (FISST) method which allows one to estimate the effect of a range of applied forces from a single molecular dynamics simulation, and also demonstrated that FISST additionally accelerates sampling of a molecule's conformational landscape.
For some problems, we find that this acceleration is not sufficient to capture all relevant conformational fluctuations, and hence here we demonstrate that FISST can be combined with either temperature replica exchange or solute tempering approaches to produce a hybrid method that enables more robust prediction of the effect of small forces on molecular systems. 
\end{abstract}

\begin{figure*}[ht!]
\centering
\includegraphics{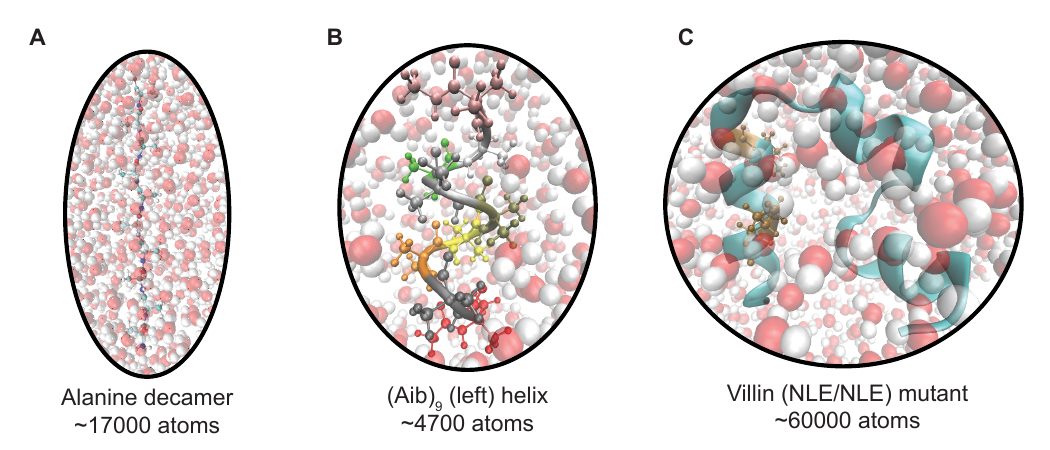}
\caption{Systems probed in this study. (A) Solvated alanine decamer starting in the extended state. (B) Solvated Aib$_9$ molecule starting from the left-handed helical state. Each residue is colored according to the residue type. (C) Solvated villin (NLE/NLE) mutant starting in the folded state. Locations of residue mutations are colored in ochre. In all cases, pulling forces are applied to the terminal C$_\alpha$ atoms.}
\label{fig:systems}
\end{figure*}
\section{Introduction}
Biological systems must have mechanisms for being able to sense and respond to mechanical forces from their environment and those that are generated internally through the action of molecular machines \cite{hoffman2011dynamic,iskratsch2014appreciating,roca2017quantifying,gomez2021molecular}. 
Cells can employ proteins to sense and respond to these forces using a wide range of molecular mechanisms which we previously reviewed \cite{gomez2021molecular}.
Perhaps the simplest such mechanism is the use of a single disordered peptide domain at the locus of a mechanical process, whose change from a collapsed to an extended conformation with single piconewtons of force could be sufficient to change the behavior of a larger protein machine.
This kind of behavior has been identified in polymerization factors called formins through a combination of in vitro and in vivo biochemistry with simple modeling \cite{courtemanche2013tension,jegou2013formin,zimmermann2017mechanoregulated,zimmermann2019feeling,lappalainen2022biochemical}, but a precise molecular mechanism for such behavior which explains the differences between homologous proteins in different species has not yet been shown \cite{zimmermann2019feeling,lappalainen2022biochemical}.

While these formin disordered domains are very large and the effect of force on their activity is complex, the effect of a pulling force on simple peptides has been exploited for the development of molecular sensors termed tension sensor modules (TSMs) \cite{freikamp2016piconewton,ham2019molecular,fischer2021molecular}.
These TSMs consist of a short protein or peptide with donor and acceptor dye molecules that can undergo fluorescence resonance energy transfer (FRET) on the termini.\cite{freikamp2016piconewton,fischer2021molecular} 
Because FRET energy transfer is highly sensitive to distance, the FRET signal can be used to infer the distance between the ends of the molecule; this distance can be converted into a force through calibration experiments performed with molecular tweezers, if a specially selected molecule is chosen which does not exhibit hysteresis.\cite{freikamp2016piconewton}
Genetically encoded TSMs can then be used to measure the forces felt by certain proteins in living cells, such as those within focal adhesion complexes, which serve as the connection between the internal cytoskeleton and the exterior environment of a cell.\cite{cost2015measure,lacroix2018tunable}
These measurements were used to confirm the relevance of 1-20 pN forces in focal adhesion behavior \cite{focal2010}.
Through experimentation, different peptides or small proteins have been found that exhibit peak force sensitivity over different ranges \cite{freikamp2016piconewton,brenner2016spider,ham2019molecular}.
This has motivated us to discover using molecular simulation approaches which properties of a disordered peptide sequence or small folded protein determine this force sensitivity, and to ask how we can design such behavior.

 Molecular Dynamics simulations (MD) can reveal highly detailed molecular-level information about a wide range of biomolecular systems  \cite{dror2012biomolecular,schlick2021biomolecular}.
To explore a biomolecule's conformational landscape using a reasonable amount of computational expense, it is necessary to employ enhanced sampling techniques that bias the system's behavior in such a way that it can more readily cross barriers in its free energy landscape \cite{tuckerman2023statistical,frenkel2023understanding}.
To do so, a wide range of techniques have been developed, most of which can be categorized by either heating part or all of the system, or adding a bias potential along some or many coordinates termed collective variables (CVs) \cite{henin2022enhanced}.
MD simulations combined with enhanced sampling techniques can be used to explore the behavior of a system experiencing a constant or time varying mechanical force \cite{gomez2021molecular,stirnemann2022recent}.
In much of our work, we have focused on the \textit{constant force} paradigm, in which case a force applied along a CV such as the end-end distance of a protein produces a simple modification to the system's Hamiltonian,
\begin{equation}
    H(\vec{q},F)=H(\vec{q})-F Q(\vec{q}),
\end{equation}
where $Q(\vec{q})$ is a CV that depends on $\vec{q}$, the configurational degrees of freedom of the system.
The negative sign convention is taken such that a positive $F$ corresponds to a pulling force, i.e. which promotes an increase in $Q$.

Motivated by the problem of computing the force-extension behavior of peptides such as disordered formin domains or peptide tension sensors, we previously developed the method Infinite Switch Simulated Tempering in Force (FISST) \cite{hartmann2020infinite}.
There, we demonstrated that it is possible to sample the effect of a range of forces on a system using a single simulation which includes a combination of (a) a special CV-dependent force, and (b) an observable weight function that allows one to reweight samples to any intermediate force, as described in the next section. 
We also demonstrated that  FISST  can promote transitions between otherwise kinetically inaccessible states of a system due to the action of the additional bias potential. 
This method was implemented and released as a module in the PLUMED open source sampling library \cite{tribello2014plumed,bonomi2019promoting}, and we also described its use in a PLUMED masterclass.\footnote{Number 22-15, \url{https://www.plumed.org/masterclass}}

However, in some cases, we find that when applying FISST to peptides or proteins for which small forces should result in a population of extended states, the system remains trapped near its initial configuration. 
We therefore wish to combine the efficiency of FISST for sampling many simultaneous forces with a method that is more effective at exploring conformational states of the molecule.

Here, we demonstrate that the performance of FISST can be improved by coupling it with Replica Exchange (RX) approaches\cite{swendsen1986replica,tuckerman2023statistical,frenkel2023understanding} using three benchmark systems of increasing difficulty (Fig.~\ref{fig:systems}).
After giving a theoretical overview of FISST and how it is naturally coupled with RX, we demonstrate that FISST combined with temperature replica exchange accelerates sampling for our previous test case of an alanine decamer\cite{hartmann2020infinite}.
We then give the example of the achiral Aib$_9$ helical peptide, where FISST alone is not enough to destabilize the folded state, but FISST combined with temperature or solute tempering allows robust sampling of the $F=0$ free energy landscape, and prediction of the force extension curve for this molecule.
Finally, we show preliminary data computing the force-extension behavior for a more complicated molecule, a villin headpiece mutant; this system is both well characterized in MD simulations and is a variant of a protein whose force-extension behavior has been measured experimentally as a TSM.\cite{austen2015extracellular,freikamp2016piconewton,ham2019molecular,fischer2021molecular}
\section{Theory}
\subsection{FISST overview}
The aim of FISST is to compute averages of observables $O(\vec{q})$ when a constant force $F$ is applied along a collective variable $Q(\vec{q})$. 
At constant temperature, this corresponds to 
\begin{equation}
    \langle O \rangle_F = \frac{\int d\vec{q} O(\vec{q}) e^{-\beta U(\vec{q}) +\beta F  Q(\vec{q})}}{Z_q(F)},
\end{equation}
where $\beta=1/(k_\mathrm{B} T)$, $U$ is the potential energy function for the system, and $Z_q(F)\equiv\int d\vec{q}e^{-\beta U(\vec{q}) +\beta F Q(\vec{q})}$ is the configurational partition function for a given $F$.

In Ref.~\citenum{hartmann2020infinite}, we showed that averages of this type can be obtained from a single simulation with a modified applied force $\bar{F}(Q)$ that is derived from the infinitely fast switching limit which would arise if sampling a ladder of applied forces from $F_\mathrm{min}$ to $F_\mathrm{max}$.

In this limit, the probability density that would be sampled is a weighted average over all forces, with weights $\omega(F)$ that say how important each force is,
\begin{equation}
    \bar{\rho}(\vec{q})=\frac{\int_\fmin^\fmax dF' \omega(F') e^{-\beta U(\vec{q}) +\beta F'  Q(\vec{q})} }{\int_\fmin^\fmax dF' Z_q(F') \omega(F') }
\end{equation}

From this distribution, we get the potential of mean force up to an additive constant through
\begin{equation}
    e^{-\beta A(\vec{q})} \equiv \int_\fmin^\fmax dF' \omega(F') e^{-\beta U(\vec{q}) +\beta F'  Q(\vec{q})} 
    \label{eq:pmf}
\end{equation}
The FISST algorithm attempts to learn $\omega(F)$ ``on-the-fly'' such that forces are sampled evenly, which occurs when $\omega(F)\propto 1/Z_q(F)$, and this is accomplished in an iterative manner.
After doing so, the integral of their product becomes a constant $C\equiv\int_{\fmin}^\fmax dF' \omega(F')Z_q(F')$

From the potential $A(\vec{q})$ in Eq.~\ref{eq:pmf}, we can get the forces to apply in an MD simulation that will sample from this probability density by taking the negative gradient with respect to atomic positions,
\begin{equation}
    -\nabla A(\vec{q}) = -\nabla U + \bar{F}(Q) \nabla Q,
    \label{eq:newforce}
\end{equation}
where
\begin{equation}
    \bar{F}(Q) = \frac{\int_\fmin^\fmax dF' \omega(F') F' e^{\beta F'  Q(\vec{q})} }{\int_\fmin^\fmax dF' \omega(F') e^{\beta F'  Q(\vec{q})} }.
\end{equation}
The FISST module in PLUMED works by computing $\bar{F}(Q)$ for any choice of CV $Q$, and then modifying the forces used in any compatible MD engine by adding $\bar{F}(Q) \nabla Q$.
We note that this need not be a simple force/distance pair, but could be a more general quantity, e.g. a tension and an area or an electric field and a dipole moment.

After simulating with this modified potential, averages of observables at different forces can be computed from a weighted average over $N_t$ snapshots by including `observable weights' $W_F(\vec{q})$ computed on the fly\cite{hartmann2020infinite},
\begin{equation}
    \langle O \rangle_F = \frac{1}{N_t}\sum_{i=1}^{N_t} W_F(\vec{q}(t)) O(\vec{q}(t)),
\end{equation}
where
\begin{equation}
    W_F(\vec{q})=\frac{C e^{\beta U(\vec{q})+\beta F Q(\vec{q})}}{Z_q(F) \int dF' \omega(F')e^{\beta (F'-F)Q(\vec{q})}}
    \label{eq:obsv_weights}
\end{equation}

\subsection{Combination of FISST with replica exchange}
In replica exchange simulations, a Markov Chain Monte Carlo procedure is carried out, with detailed balance in exchanges ensuring that each replica maintains a particular equilibrium distribution \cite{tuckerman2023statistical,frenkel2023understanding}. 

In Hamiltonian replica exchange, each replica is simulated via its own Hamiltonian $H_i$, which could be simulated at inverse temperature $\beta_i$.
Within each copy of the simulation, configurations appear with probability $P_i(\vec{q}) \propto \exp(-\beta_i H_i(\vec{q}))$ \cite{fukunishi2002hamiltonian}. 
Ensuring detailed balance of exchange between configurations $\vec{q}$ and $\vec{q'}$ generated from Hamiltonians $H_i$ and $H_j$ respectively using a Metropolis criterion requires that \cite{fukunishi2002hamiltonian,bussi2014hamiltonian}.
\begin{equation}
    P_\mathrm{accept} = \min\left(1, \frac{P_i(\vec{q'})P_j(\vec{q)}}{P_j(\vec{q'})P_i(\vec{q})}\right)
\end{equation}

In this case,
\begin{eqnarray}
\frac{P_i(\vec{q'})P_j(\vec{q)}}{P_j(\vec{q'})P_i(\vec{q})}=\frac{e^{-\beta_i H_i(\vec{q}')}e^{-\beta_j H_j(\vec{q})}}{e^{-\beta_i H_i(\vec{q})}e^{-\beta_j H_j(\vec{q}')}}\\
=e^{-\beta_i(H_i(\vec{q}')-H_i(\vec{q}))+\beta_j(H_j(\vec{q}')-H_j(\vec{q}))}
\end{eqnarray}

Below, when we combine FISST with temperature replica exchange (TRX), $\beta_i$ will be different for each replica, but we will still be performing a form of Hamiltonian exchange due to the different bias applied in each replica. 
When performing solute tempering, $\beta_i$ will be identical for all replicas, however in addition to the different bias applied, the potential energy function will also be different between replicas in such a way as to represent effectively higher solute temperatures (see below).

Hamiltonian RX is implemented in GROMACS through the PLUMED plugin library \cite{tribello2014plumed,bussi2014hamiltonian,bonomi2019promoting}, which computes the energy under both Hamiltonians by swapping positions and momenta between replicas and re-evaluating the total energy.
If different biases are applied from PLUMED, then this is also recomputed for the swapped configurations and taken into account in the acceptance criterion \cite{bussi2014hamiltonian}.

To enable the combination of FISST with RX, we modified our PLUMED implementation such that statistics gathered for computing the quantities $\omega_i(F)$ and  $Z^i_q(F)$ which are needed for computing the on-the-fly force $\bar{F}_i(Q)$ are properly computed during the exchange procedure (prior code would update statistics every time the bias is computed, which occurs three times during the exchange procedure).
This revised code is available from the github page for this paper (see data availability statement) and will be contributed to our FISST module in the public PLUMED library soon.

Computing these quantities using data from the parallel simulations should improve convergence of the weights, however as discussed in \cite{martinsson2019simulated,hartmann2020infinite} the observable weights are correct even before these quantities are converged, and in practice the weights assigned to each force, $\omega(F)$, converge quickly so this does not have a major effect. 
We also perform simulations where the weights are fixed after an initial equilibration phase.

The partial tempering variants that we have tested are based on the Replica Exchange with Solute Tempering (REST) idea \cite{liu2005replica,huang2007replica,wang2011replica}. In this work, we use REST3 \cite{zhang2023re} which scaled interactions between solute and solvent in a way that was shown to not suppress extended configurations of peptides at higher effective temperatures as could occur with earlier REST variants.
REST2 and REST3 are formulated such that for $N$ replicas, the target temperature of replica $i$ is given by $T_i=T_0 (\frac{T_\mathrm{max}}{T_0})^{i/(N-1)}$, for $i$ from 0 to $N-1$ \cite{wang2011replica,zhang2023re}. The potential energy function of each replica $U_i(\vec{q})$ scales the protein-protein and protein-water interactions by factors $\lambda_i^{pp}$ and $\lambda_i^{pw}$ respectively \cite{wang2011replica,zhang2023re}. REST3 introduces an additional scaling factor for the non-electrostatic contributions to the protein-water interactions $\kappa_i$. The total potential energy in replica $i$ is then given by,
\begin{align}
    U_i(\vec{q}) &= \lambda_i^{pp} U_{pp}(\vec{q}) + \lambda_i^{pw} U^{elec}_{pw}(\vec{q}) \\ \nonumber
    &+ \kappa_i \lambda_i^{pw} U^{non-elec}_{pw}(\vec{q}) + U_{ww}(\vec{q}),
\end{align}
with $\lambda_i^{pp}=T_0/T_i$, $\lambda_i^{pw}=\sqrt{T_0/T_i}$, and $\kappa_i=1+0.005(m-3)(m>3)$, with REST2 being recovered if $\kappa_i$ is set to unity for all $i$ \cite{wang2011replica,zhang2023re}.

\section{Results and Discussion}
\subsection{Description of systems}
For this study, we evaluated the FISST+RX approach on the three atomistic systems shown in Fig.~\ref{fig:systems}: an alanine decamer Ala$_{10}$, an achiral peptidic helix of $\alpha$-aminoisobutyric acid Aib$_9$ \cite{botan2007energy,buchenberg2015hierarchical}, and the fast folding Villin Norleucine double mutant (HP35-NLE/NLE). 
Further details for system construction, equilibration, and production runs can be found in the Sec.~\ref{sec:methods}. 
In brief, we used the same setup for Alanine decamer (Fig.~\ref{fig:systems}A) as our original work\cite{hartmann2020infinite}. 
Alanine 10 was parameterized in CHARMM36\cite{mackerell2012charmm36} forcefield, solvated in explicit TIP3P water, and neutralized with no additional ions added. 
Input files for Aib$_9$ (Fig.~\ref{fig:systems}B) were provided by the authors of Ref.~\citenum{mehdi2022accelerating}. 
As in Ref.~\citenum{mehdi2022accelerating} and our previous study\cite{sasmal2023reaction}, the Aib$_9$ helix starts in the left handed state, parameterized in CHARMM36m\cite{huang2017charmm36m} forcefield, solvated in TIP3P, and neutralized with no additional ions added.
Finally, we used the same setup for Villin (NLE/NLE) mutant (Fig.~\ref{fig:systems}C) as described in Ref.~\citenum{sasmal2023reaction} chosen to match Ref.~\cite{piana2012protein}.  Villin mutant was solvated in TIP3P, parameterized in Amberff99SB*-ILDN\cite{lindorff2010improved} forcefield, and neutralized and ionized to a 40 mM salt concentration. The final sizes of each system were roughly 17000, 4700, and 60000 atoms respectively. Sodium ($\ce{Na+}$) and ($\ce{Cl-}$) ions were used for neutralization and ionization. 

\subsection{Simulations of polyalanine validate implementation of hybrid sampling approach}
In our previous work, we demonstrated using the alanine decamer (Ala$_{10}$) that FISST could accurately compute the end-end distribution at a range of forces from a single simulation, as compared to a reference TRX simulation.\cite{hartmann2020infinite} Taking this as a stand-in for the more complicated peptides that we wish to probe in the future, we chose this as a benchmark to check that combining FISST with RX does not degrade performance. 

As in our previous work, we began by computing the end-to-end distance probability distribution functions for forces ranging from $-10$~pN to 10~pN.
We compare our work to the same FISST and TRX data generated for Ref.~\citenum{hartmann2020infinite}, which consists of a 500 ns FISST run at $T=300K$ using a force range [-10pN:10pN], and TRX consisting of 40 replicas run for 160 ns each (6.4 $\mu$s total simulation time) with temperatures ranging from 300K to 400K.

For this study, we first combined TRX and FISST by running 40 parallel FISST simulations at the same temperatures as our reference TRX simulation, using 100 ns for each replica and a force range of [-10pN:10pN].
We also ran simulations with the same FISST parameters using FISST+REST3 on 10 replicas spanning 300K to 600K for  400 ns each. 
As a control, we also ran REST3 simulations with constant forces of -10, 0, and 10 pN forces for the same solute temperature ranges, number of replicas, and simulation times. 
For all analyses presented here, we compute results using the bottom replica.  

\begin{figure*}[ht!]
\centering
\includegraphics[width=0.75\textwidth]{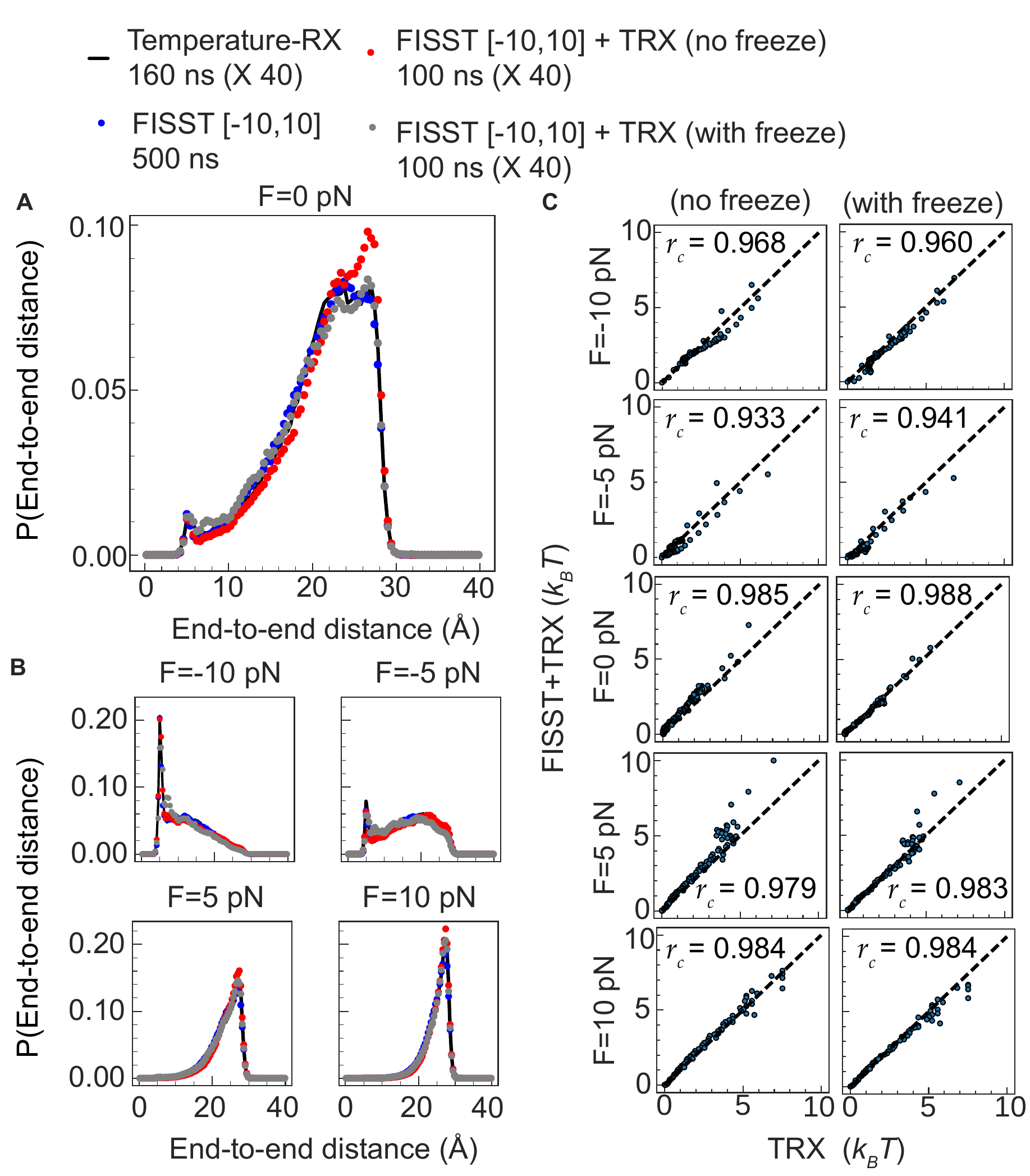}
\caption{(A) End-end distance distributions for Ala$_{10}$ at $F=0$ for TRX (black solid line), FISST (blue spheres), FISST+TRX without freezing weights (red spheres), and FISST+TRX with freezing (gray spheres). (B) End-end distributions for $F=$-10, -5, 5, and 10 pN. (C) Comparison of free energies (see Fig.~\ref{fig:ala10figS1}) computed from end-end distribution functions, comparing FISST+TRX with and without freezing to corresponding reference TRX data at -10, -5, 0, 5, 10 pN forces.}
\label{fig:ala10fig1}
\end{figure*}
In Fig.~\ref{fig:ala10fig1}, we show a comparison of these methods for fixed forces of -10, -5, 0, 5, and 10 pN.
We find a reasonable visual agreement from all our simulation methods relative to our previous FISST and Temperature replica exchange data at all forces, including a peak representing a collapsed state at $\sim$5 $\textup{\AA}$ for 0, -5, and -10 pN (Fig.~\ref{fig:ala10fig1}A,B).

However, our initial FISST+TRX run for which we did not freeze the FISST weights shows a slightly higher peak at $\sim$27 $\textup{\AA}$ compared to the FISST and Temperature RX runs at the zero force (Fig.~\ref{fig:ala10fig1}A, red spheres). 
We repeated these runs with frozen weights obtained by simulating the parallel replicas without any exchange attempts for 20 ns each, and then continuing with FISST+TRX with those weights fixed using the \texttt{FREEZE} option in the FISST code. In this case, the data with freezing (Fig.~\ref{fig:ala10fig1}A, gray spheres) has a more accurate peak at $\sim$27 $\textup{\AA}$ and better qualitative agreement with the TRX reference.

To check our results quantitatively, we computed the free energy profiles $A(Q)$ from the probability distribution functions at the different forces shown in Fig.~\ref{fig:ala10fig1}A,B by taking $A(Q)\equiv-k_\mathrm{B} T\ln(P(d_\mathrm{End}))$ and subtracting an offset such that the minimum in all cases was zero (see Fig.~\ref{fig:ala10figS1}).
We then constructed scatter plots of the two sets of FISST+TRX free energies (with and without freezing the weights) versus the TRX data at each of the corresponding forces, with results shown in Fig.~\ref{fig:ala10fig1}C.
For each of the free energy scatter plots
we computed the Spearman's rank correlation coefficient \cite{sammut2011encyclopedia} ($r_c$) using \texttt{stats.spearmanr} function implemented in \texttt{scipy} \cite{2020SciPy-NMeth,corder2014nonparametric}.
Although both sets of data gave relatively high $r_c$ values when averaged over the 5 forces (0.9698 and 0.9712 without and with freezing the weight distributions, respectively), we observe a slight improvement when freezing the weights for the cases of -5, 0, and 5 pN forces. 

\begin{figure*}[ht!]
\centering
\includegraphics[width=0.75\textwidth]{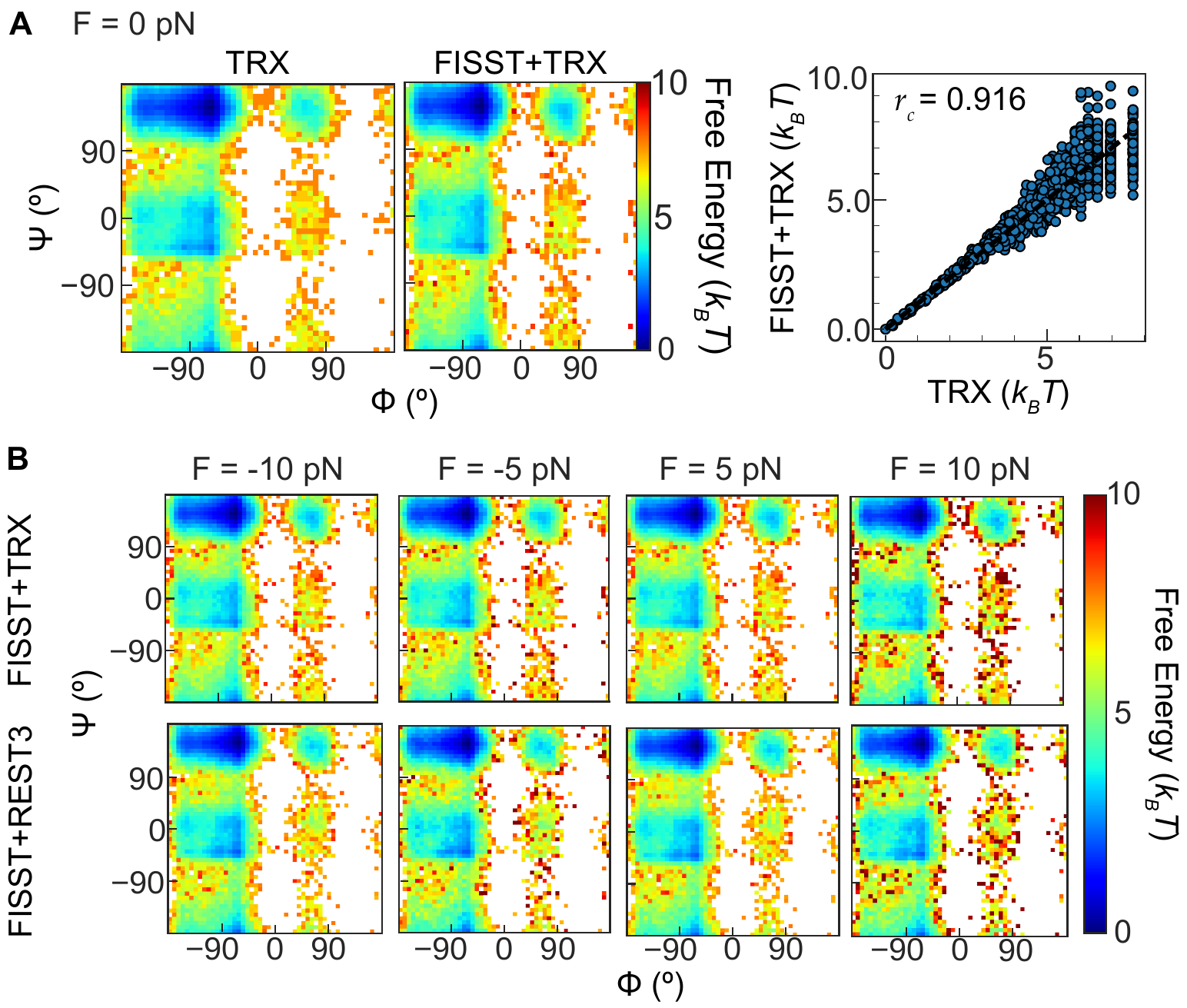}
\caption{(A) (left) TRX and FISST+TRX Ramachandran plot at zero force and (right) scatter plot comparing  FISST+TRX and TRX free energies at zero force. (B) Ramachandran plots at -10, -5, 5, and 10 pN forces for FISST+TRX (top row) and FISST+REST3 (bottom row).}
\label{fig:ala10fig2}
\end{figure*}

We emphasized in our previous work\cite{hartmann2020infinite} that using the observable weights (Eq.~\ref{eq:obsv_weights}) we are able to reconstruct averages of other observables at any force besides the one that was biased. 
In addition to end-to-end distance, we also reconstructed Ramachandran plots for Ala$_10$ FISST+TRX simulations, reweighing at the zero force shown in Fig.~\ref{fig:ala10fig2}A. 
We also computed $r_c$ between the free energies in both sets of data and found a relatively high value of 0.916. 
In Fig.~\ref{fig:ala10fig2}B we show the result of reweighting the backbone dihedral angles at -10, -5, 5, and 10 pN forces and noted the expected strengthening of the PPII basin (top left) at high force and destabilization of the alpha-helical basin right center as force is increased \cite{stirnemann2013elasticity}.    

To demonstrate the accuracy and generality of our implementation, we also performed FISST simulations coupled to REST3 and repeated our End-to-end distance and Ramachandran angle analysis for the -10, 0, and 10 pN forces, using REST3 simulation data collected at those forces. 
We carried an initial FISST+REST3 run without freezing the weights and another FISST+REST3 by freezing the weights after 20 ns in an analogous fashion, and compared the results of each run to our reference data (Fig.~\ref{fig:ala10figS2}, Fig.~\ref{fig:ala10figS3}). 
In Fig.~\ref{fig:ala10figS2}A we again find reasonable qualitative agreement in the probability distribution functions at all the forces shown. 
Qualitative analysis in Fig.~\ref{fig:ala10figS2}A finds relatively high $r_c$ values for both FISST+REST3 runs, an average of 0.982 and 0.977 when freezing the weights and without, indicating a slight improvement with respect to the reference REST3 calculations when freezing. 
We also reweighted the Ramachandran angles calculated from FISST+REST3 (Fig.~\ref{fig:ala10fig2}B (bottom row)) at the forces shown to visually demonstrate that the combination of FISST+REST3 gives equivalent results to FISST+TRX, with quantitative analysis shown in Fig.~\ref{fig:ala10figS3}. 

\subsection{Simulations of Aib$_9$ show improved performance from hybrid FISST+RX sampling}
While our results on Ala$_{10}$ show that we are able to combine FISST with replica exchange techniques, they do not demonstrate an obvious improvement that requires such a hybrid method. 
In this section, we show that FISST alone may not be able to sample the free energy landscape of a structured peptide, necessitating the additional sampling from tempering.

 Here we analyze results for the achiral Aib$_9$ system starting from a left-handed configuration at $T=400K$, for which we initially performed a 4.0 $\mu$s unbiased simulation and a 2.0 $\mu$s FISST simulation for the force range [-10pN:20pN].
 While the unbiased simulation shows a transition rate of approximately 1 inversion per 2 $\mu$s (Fig.~\ref{fig:AIB9figS1}), the FISST simulation actually does not, showing a case where FISST can impede conformational exploration. Hence we felt this is an ideal test system for demonstrating the effectiveness of FISST+RX.

\begin{figure}[ht!]
\centering
\includegraphics[width=0.75\columnwidth]{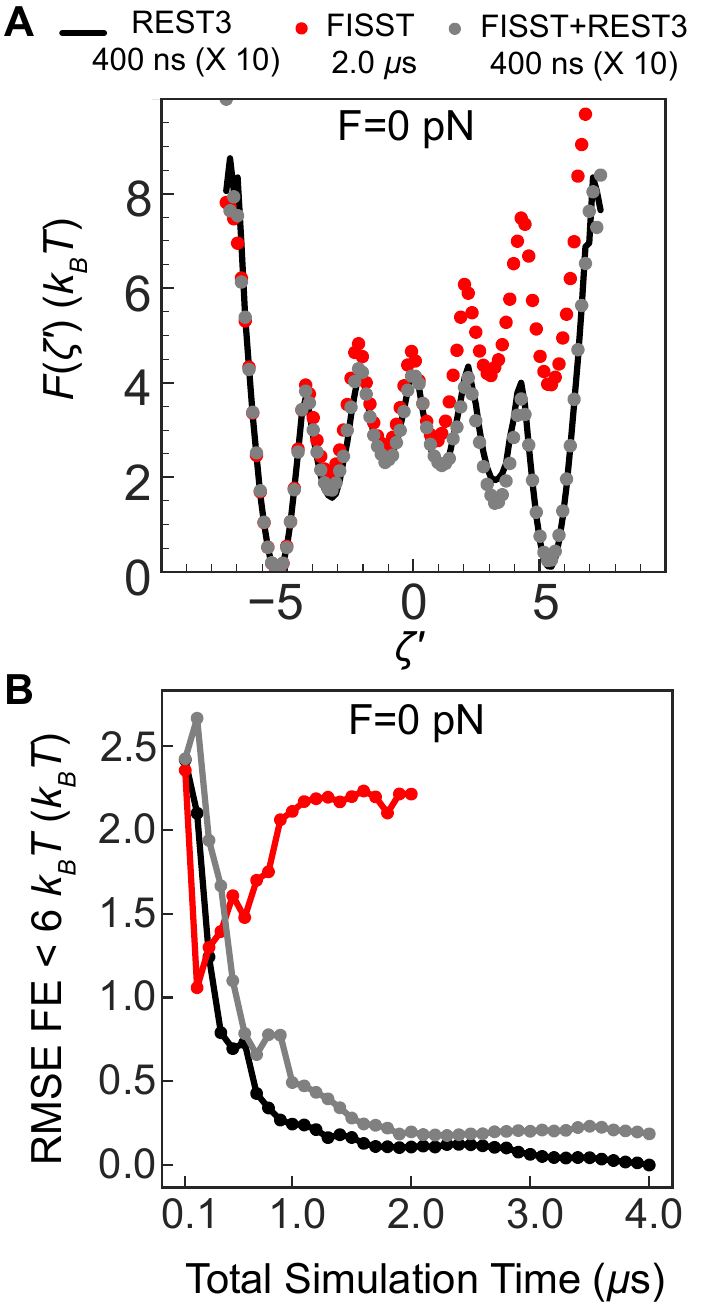}
\caption{(A) $F(\zeta')$ at zero force calculated from REST3 (black solid line), FISST (red spheres), and FISST+REST3 (gray spheres). (B) RMSE of $F(\zeta')$ for values below 6 $k_\mathrm{B}T$ using data points from different simulation time windows.}
\label{fig:AIB9fig1}
\end{figure}

We measured the chirality transition of the Aib$_9$ helix using the $\zeta'$ coordinate defined as the negative sum of the five inner $\phi$ dihedral angles shown in Fig.~\ref{fig:AIB9figS1}, consistent with previous studies.\cite{buchenberg2015hierarchical,mehdi2022accelerating,sasmal2023reaction}
With this definition using angles in radians, the left handed helical configuration takes on a value of $\zeta'=-5$ and the right handed $\zeta'=5$.
We constructed $F(\zeta')$ at zero force for each of our simulations using 100 equally spaced windows starting from $\zeta'=-7.5$ to $\zeta'=+7.5$ (Fig.~\ref{fig:AIB9fig1}A).
When performing 4$\mu$s (400 ns $\times$ 10 replicas) of REST3 simulations spanning 400K--800K, a symmetric free energy profile is obtained (black solid line) from the lowest replica in the ladder, showing that solute tempering is an effective sampling approach for this model problem.  
This is in contrast to the FISST data (red spheres) which fails to sample the right-helix basin even after 2 $\mu$s of simulation time. 
 
 We then proceeded to combine FISST with REST3 by using the same 10 replicas but adding FISST sampling on the range [-10pN:20pN].  
 We find that the FISST+REST3 data (gray spheres) not only overcomes the poor sampling from the FISST method alone but also converges with the benchmark REST3 data. Snapshots depicting some molecular configurations observed in this process are shown in Fig.~\ref{fig:AIB9figS2}.
To quantify the accuracy of the combined sampling, in 
Fig.~\Ref{fig:AIB9fig1}B we computed the root-mean-squared error (RMSE) of $F(\zeta')$ for free energies below  $6 k_\mathrm{B}T$ (chosen to encompass all of the metastable states based on Fig.~\ref{fig:AIB9fig1}A). We computed the RMSE at $F=0$ for progressively longer time windows starting with 100 ns.
At short times, the simulation does not adequately sample the entire $\zeta'=[-7.5,7.5]$ range as it remains near the left-handed state, resulting in a high error.
While as previously noted, the FISST alone simulation does not converge, the FISST+REST3 converges towards the REST3 reference to less 0.5 kcal/mol ($\sim$0.63 $k_\mathrm{B} T$ for $T=400$K) in approximately 300 ns of total sampling.
The same trends hold when using all bins for the RMSE calculation (Fig.~\ref{fig:AIB9figS3}).
While the FISST+REST3 curve does not approach zero, this appears to be due to simply finite sampling resulting in slightly more data in the almost-right metastable state for FISST+REST3 and slightly more data in the almost-left state in the reference calculation.

 It should also be emphasized that from a practical point of view, if many processors are available, the FISST+REST3 may be faster in wall clock time than running a single long trajectory, where for example we needed a microsecond or more of Metadynamics simulation to converge a good free energy profile for this system, even with a good reaction coordinate \cite{sasmal2023reaction} (see Table~\ref{tab:AIB9comptime} for simulation times). 
We also note that the FISST+REST3 data also contains additional information about all forces from -10 to 20 pN, which we will discuss next, making it much more efficient when this data is needed.

\begin{figure}[ht!]
\centering
\includegraphics[width=0.75\columnwidth]{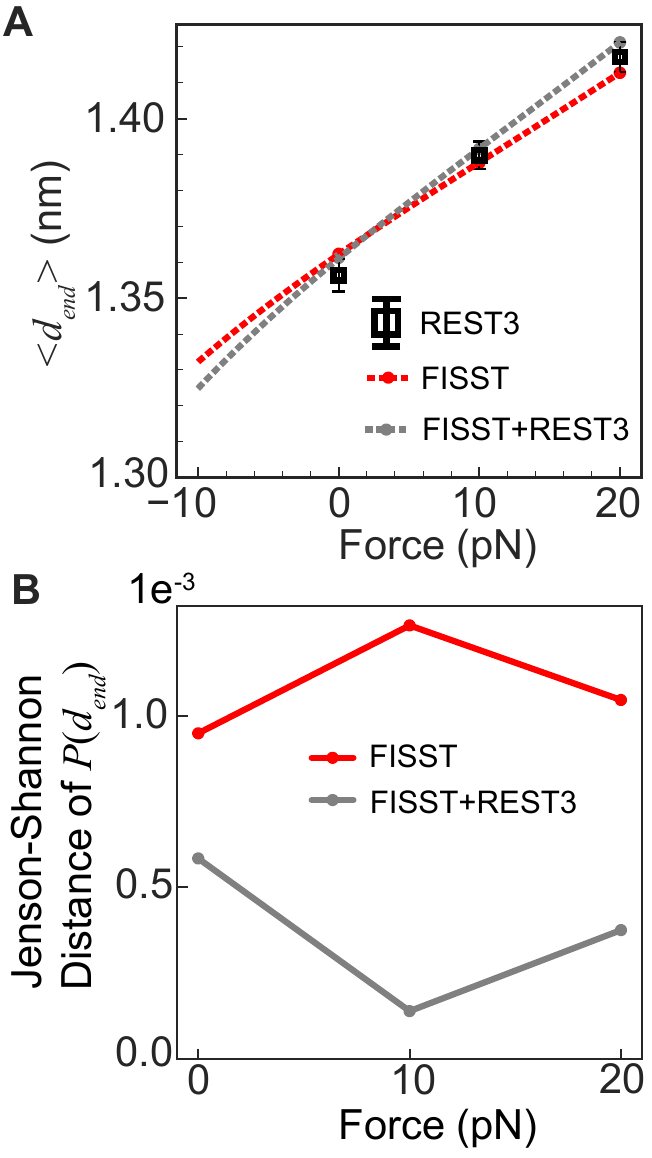}
\caption{(A) Theoretical Force vs average End-to-end distance $\langle d_{End} \rangle$ curve for Aib$_9$ calculated from FISST (red dashed line) and FISST+REST3 (gray dashed line). End-to-end distance values calculated from REST3 simulations at 0, 10, and 20 pN forces (black), and unbiased MD (blue) are embedded for comparison. (B) Jenson-Shannon distance of $P(d_{End})$ calculated for FISST (red) and FISST+REST3 (gray) for 0, 10, 20 pN forces using REST3 simulation data as the reference.}
\label{fig:AIB9fig2}
\end{figure}

In Fig.~\ref{fig:AIB9fig2}A we show the force-extension curve obtained from our simulations, by reweighting the End-to-end distance data to compute a mean distance $\langle d_\mathrm{end}\rangle$ as a function of force. 
This is an academic exercise since experimental data for this system is not available. 
Here we compare the force-extension curve for reference REST3 simulations performed at different forces of 0, 10, and 20 pN with our FISST alone or FISST+REST3 simulations.
We assess accuracy in Fig.~\ref{fig:AIB9fig2}B as we did in Ref.~\citenum{hartmann2020infinite} by computing the Jenson-Shannon Distance\cite{schindelin2003jsd} between the reweighted End-to-end probability distributions.
While this analysis shows that our FISST+REST3 result is accurate, it also appears that the FISST alone result is accurate. 
This is because, evidently, the Aib$_9$ helix is quite resistant to extensional force, and the response is predicted correctly even when trapped in only one helical state.

\subsection{Villin (NLE/NLE) mutant simulations allow us to assess performance on a TSM-like molecule}
The resistance of Aib$_9$ to pulling prevents us from showing the full extent of FISST+REST3's performance on force-extension curves. 
We now wanted to test our approach for a protein used in a TSM. 
Many such peptides do not have known structures (because they are not ordered), making it difficult to know if we are using a good starting structure or forcefield.
We therefore decided to study the villin headpiece domain (HP35) since this protein is both well characterized in experiment and probed as a tension sensor module \cite{austen2015extracellular}.
Initial test simulations we performed using wild type HP35 showed little stretching within available simulation time, even with parallel tempering approaches, which we attributed to potential forcefield over-stabilization of collapsed states \cite{piana2015water,rauscher2015structural}; forcefield choice has also been shown to have a very strong effect on the predicted stability of villin in solution \cite{robustelli2018developing}.
For this work, we therefore elected to study the Villin (NLE/NLE) mutant whose behavior has been extensively characterized and studied across many simulation studies, and in particular was exhaustively sampled by the DE Shaw Research group \cite{piana2012protein}.

We analyze data collected from two sets of REST3 simulations consisting of 8 replicas, one with a solute temperature range from 298K to 450K and another ranging from 360K to 500K. For each solute temperature range, we ran  FISST+REST3 using a force range [-10pN:20pN]  for $\sim$200 ns each (1.6 $\mu$s total simulation time).
We then repeated these simulations restarting from the point of 20 ns of simulation with weights frozen. 
We also collected data for $F=10$ and 20 pN for both solute temperature ranges. An additional reference that we include in our analysis is the $\sim$310 $\mu$s unbiased simulation of Villin (NLE/NLE) at 360K Ref.~\citenum{piana2012protein}. 
Consistent with analysis of our other systems, we analyze only the bottom replica. 

\begin{figure}[ht!]
\centering
\includegraphics[width=0.75\columnwidth]{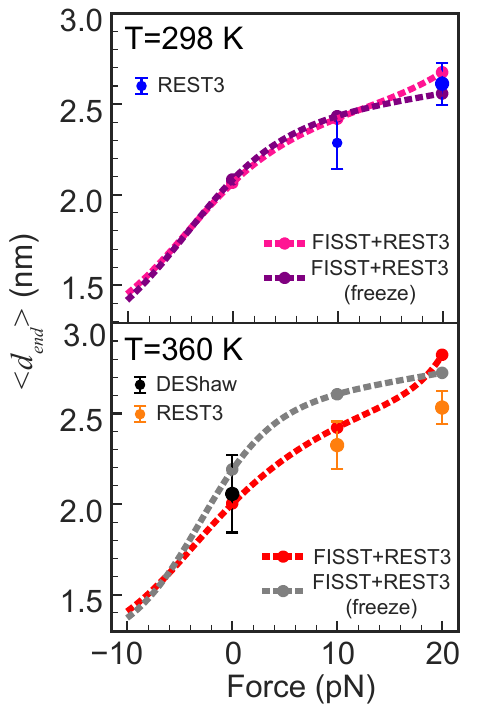}
\caption{(Top) Villin Mutant Force vs average End-to-end distance $\langle d_{End}\rangle$ calculated for FISST+REST3 without freezing the weights and freezing the weights at 298K. $\langle d_{End}\rangle$ values calculated from REST3 simulations at $F=10$ and 20 pN. Error bars represent 1/3 of the standard deviation in length at that force. (bottom).
 Similar data as above for simulations where the lowest replica is at $T=360K$, which is close to the melting temperature. Also shown is data from a 310 $\mu$s trajectory at $T=360K$ from Ref.~\citenum{piana2012protein}.}
\label{fig:hp35fig1}
\end{figure}

 Fig.~\ref{fig:hp35fig1} shows our computed force extension curves for the two temperatures selected. The lower is room temperature, where single molecule pulling experiments are performed on TSMs, and 360K is close to but below the melting temperature for the mutant using this forcefield, so that many more folding/unfolding events are observed in long unbiased simulations \cite{piana2012protein}.
 
Here we observe an elastic regime where the force extension curve is linear from around $-5$~pN to $+5$~pN of pulling force at both temperatures. 
In both cases, there is a turnover to an inextensible regime, although at each temperature, one of the two data sets shows an indication of entering another stretching regime. 
At $T=360K$, the FISST+REST3 curves lie above the reference calculations, which may be an indication of additional sampling of unlikely extended states due to additional sampling from using the hybrid method. 
We argue that this is due to enhanced sampling rather than hindered sampling, because if anything we would naively expect the FISST method to promote spending times at smaller extensions due to the need to sample the full force range from -10 to 20 pN, but the opposite is observed here.

In Fig.~\ref{fig:hp35figS1} we show how the histograms are transformed as the force on the ends of villin is increased. For both 298K and 360K, there is a prominent peak at shorter distances ($\sim 1.2$~nm) for low force which is shifted to a prominent peak at larger lengths ($\sim 2.5$~nm).
The high force distributions are unimodal, although there is some evidence for a shoulder developing at 3.0~nm at the highest forces.
The linear increase in average length due to a shift between two populations is something we discussed as the most likely scenario for the low force regime when there are two possible states \cite{gomez2021molecular}, however here we are still remaining within compact states, meaning that predominantly unfolded states are not being accessed here.
In contrast, experimental data on wild type HP35 shows a full unfolding with a change in length of 7 nm over this force range.
For this situation, we previously speculated based on geometric arguments that the force extension curve had an initial behavior like we observe here up through $\sim 5$ pN, followed by a separation of the folded state into three independent helices up to around 10 pN, at which point the helices begin to populate fully extended states \cite{gomez2021molecular}, commensurate with the discussion on folded TSMs in Ref.~\cite{ham2019molecular}.
We hypothesize that our lack of observation of this behavior in the experimentally probed force regime still corresponds to over-stabilization of the folded state or collapsed partially unfolded states by the forcefield/water model.

\section{Conclusions}
In this work, we demonstrated that our force tempering method can be enhanced through combination with replica exchange approaches. 
Combination with solute tempering showed definitive improvement for a test case where FISST alone failed. 
The combined approach is much more efficient than running many individual simulations at different fixed forces when attempting to compute a full force-extension curve as in our final example of the HP35 protein.
Also, when combining FISST with TRX, the full force extension profile at all temperatures is obtained simultaneously.

We chose to combine force and temperature sampling by employing a replica exchange approach, which we did because the implementation via a Monte Carlo scheme was practically realizable due to the efforts of the developers of PLUMED and GROMACS \cite{bussi2014hamiltonian,abraham2015GROMACS}.
However, we would also like to note that it should be possible to combine infinite switch simulated tempering in force with the infinite switch simulated tempering in temperature (ISST), upon which FISST was originally based \cite{martinsson2019simulated}.
This may be more effective than our approach here, since at least on paper the infinite switch limit is the most efficient choice for parallel tempering \cite{sindhikara2008exchange,sindhikara2010exchange,martinsson2019simulated}.
While ISST is implemented in the MIST library \cite{bethune2019mist}, combining the two approaches would require efficient implementation of estimating partition functions and weights using two dimensional integrals over both inverse temperature and force which could pose a numerical challenge, hence we chose not to pursue that effort at this time.

Finally, even with our improved sampling method, we have not computed a force extension curve that yet matches one measured experimentally.  
While it is possible that the difference is due to the difference in solvent conditions (experiments mostly performed in phosphate-buffered saline or similar), or that the experimental curve is not quite right, given the complex setup using tethering molecules and the need to significantly postprocess data from many pulling runs \cite{austen2015extracellular,ham2019molecular}, for now we presume that the larger error comes from the simulation side.
Given that we have implemented and then improved an effective force sampling approach, this points us towards considering alternative water models and protein forcefields (or modifying terms in the current ones) to find one that best matches the known behavior for a molecule like villin. We hope that combining our effective sampling approach with the appropriate forcefield will allow us to design \textit{in silico} new tension-sensing peptide molecules.

\section{Methods}\label{sec:methods}
\subsection{System details}
\noindent \textbf{Ala$_{10}$}---This system is the same as that used in our previous FISST study \cite{hartmann2020infinite}.
In summary, the system consists of a cubic box of size 56.0~\textup{~\AA}, solvated using TIP3P water \cite{jorgensen1983comparison} and parameterized using the CHARMM36 forcefield. 
The total system size is 17293, including 5730 water molecules. The system is simulated at 300K. 

\noindent
\textbf{Aib$_9$}---GROMACS \cite{abraham2015GROMACS} inputs were provided by the authors of Ref.~\citenum{mehdi2022accelerating}. 
The system consists of a cubic box of length 35.0~\textup{~\AA}, solvated using TIP3P water molecules \cite{jorgensen1983comparison}, and parameterized using the CHARMM36m forcefield \cite{huang2017charmm36m}. The total size of the system was 4749 atoms including 1540 water molecules. The net charge of the system was neutral with no additional ions added. 
The system is simulated at 400K.

\noindent
\textbf{Villin Mutant}---Inputs for this system were those generated according to the protocol in Ref.~\citenum{sasmal2023reaction}. 
The Villin mutant (PDB ID: 2F4K) was constructed in a cubic box of length 86.80~\textup{~\AA}, solvated using TIP3P water molecules \cite{jorgensen1983comparison}, and parameterized using Amberff99SB*-ILDN forcefield \cite{lindorff2010improved}. 
The total system size is 60392 atoms including 19928 water molecules. The system was neutralized and ions were added to bring the system to a 40 mM salt concentration (15 $\ce{Na+}$ ions, 16 $\ce{Cl-}$ ions). 
The system is simulated at 298K and 360K.

\subsection{Production runs} 
\subsubsection{Overview}
Production data were collected using the GROMACS MD engine \cite{abraham2015GROMACS}. 
All single process MD were run in GROMACS 2020.4, while Hamiltonian exchange simulations were run in GROMACS 2019.6 patched with PLUMED version 2.7.0 \cite{bonomi2019promoting}.
Simulations performed at constant force employed the \texttt{RESTRAINT} feature in PLUMED \cite{tribello2014plumed}.

\subsubsection{FISST details}
The FISST\cite{hartmann2020infinite} algorithm and single force calculations (applied with the RESTRAINT keyword) were performed using PLUMED\cite{bonomi2019promoting}. 
In all cases, the bias is applied along a collective variable that is the distance between the first and last C$_\alpha$ atoms of the peptides. 
The FISST force range chosen for Ala$_{10}$ was [-10pN:10pN] and for all other simulations [-10pN:20pN], discretized over 121 gridpoints to perform the integrals \cite{hartmann2020infinite}.
An initially uniform distribution for the force weights was used.
For Aib$_9$, weights were updated every 500 steps (1 ps) and both observable and restart data were also saved every 500 steps. 
For Ala$_{10}$ and HP35, the weights were updated every 1000 steps (2 ps), and the observable data and restart data were also saved for the same number of steps. 

\subsubsection{REST3 Simulations}
We implemented the REST3\cite{zhang2023re} algorithm for all of our multiple process MD runs. For Ala$_{10}$  we choose tempering parameters $\lambda$ and $\kappa$ parameters using the script provided by Ref.~\citenum{zhang2023re} to simulate a solute temperature range of 300K to 600K over 10 replicas, and for Aib$_9$ we chose 400K to 800K. 
For Villin mutant we ran two sets of $\lambda$ and $\kappa$ values; with one set of 8 replicas from 298K to 450K and another set of 8 replicas from 360K to 500K. Exchanges were attempted every 5 ps. Our REST3 inputs, scripts, and instructions to set up GROMACS topologies for REST3 simulations can be found on the manuscript GitHub (see below). 

\subsection{Data Analysis}
All trajectory files were analyzed using the PLUMED driver and \texttt{mdtraj}\cite{mcgibbon2015mdtraj} in python 3.8.0 Trajectory and structure files were visualized in VMD 1.9.3\cite{humphrey14vmd}.   

\section{Data Availability}
All input files, scripts, and output files are available from  a GitHub repository for this manuscript, \url{https://github.com/hocky-research-group/FISST-RX_20203}. Any additional files will be made available upon request. 

\begin{acknowledgement}
YS and GMH were supported by the National Institutes of Health through the award R35GM138312. 
This work was supported in part through the NYU IT High Performance Computing resources, services, and staff expertise, and simulations were partially executed on resources supported by the Simons Center for Computational Physical Chemistry at NYU (Simons Foundation Grant No 839534).
We would like to thank Eric Vanden-Eijnden for many stimulating conversations during the development of FISST. 
We would also like to thank Giovanni Bussi for discussing with us the details of the GROMACS/PLUMED Hamiltonian exchange implementation. We also thank the D.E. Shaw Research for providing simulation data on the HP35 protein, and the Tiwary lab for providing their input files for Aib$_9$. 
\end{acknowledgement}

{\footnotesize \bibliography{fisst_rx}}
\onecolumn
\appendix
\section{Supporting information}
\subsection{Alanine Decamer}
\setcounter{figure}{0}
\renewcommand{\thefigure}{S\arabic{figure}}
\begin{figure}[ht!]
\centering
\includegraphics[width=0.8\textwidth]{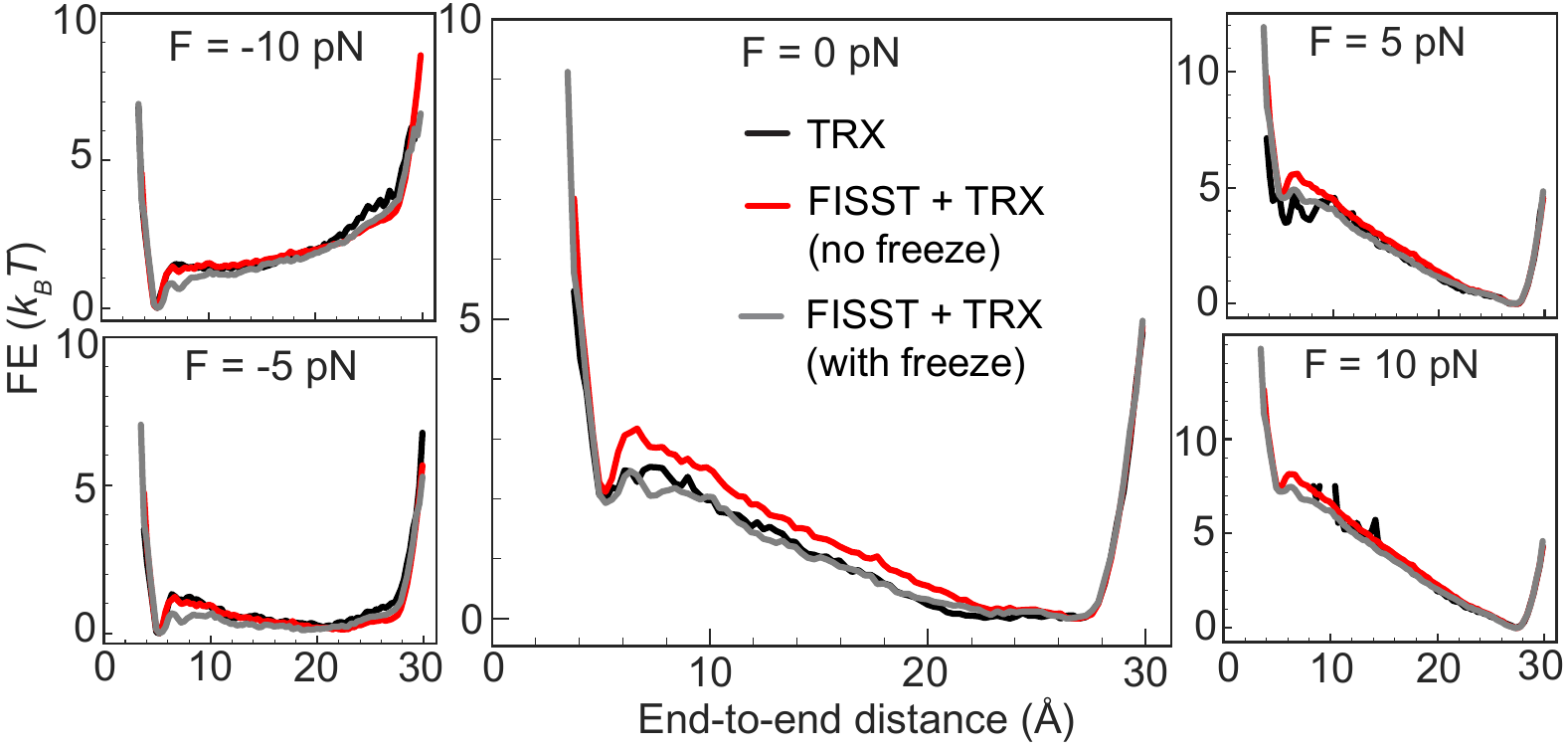}
\caption{Alanine Decamer Free energy profiles for TRX (black solid line), FISST+TRX without freezing weights (red solid line), and FISST+TRX with freezing weights (gray solid line) calculated from the End-to-end distance probability distributions at -10, -5, 0, 5, and 10 pN forces shown in Fig.~\ref{fig:ala10fig1}(A) and Fig.~\ref{fig:ala10fig1}(B).}
\label{fig:ala10figS1}
\end{figure}
\begin{figure}[ht!]
\centering
\includegraphics[width=0.75\textwidth]{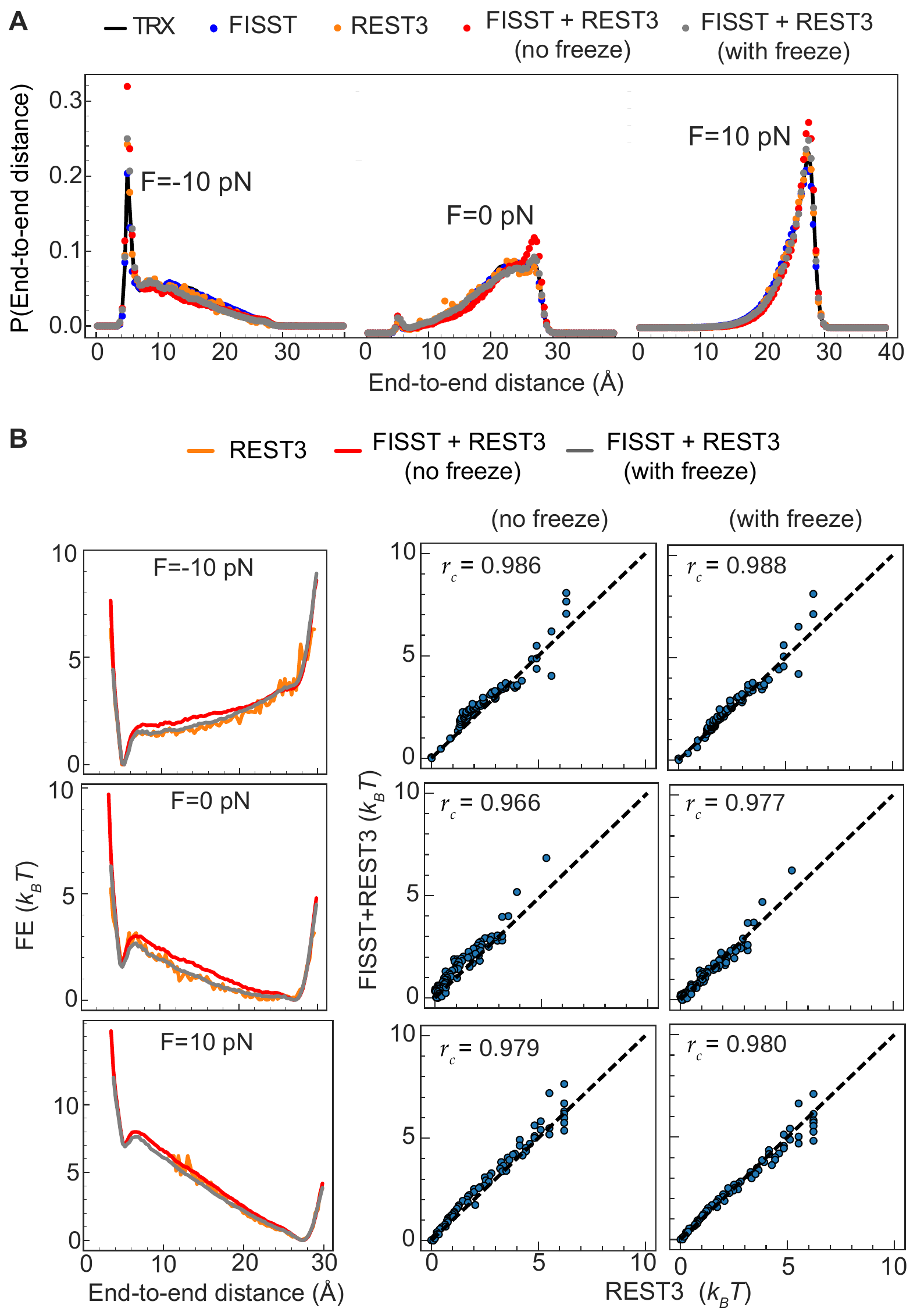}
\caption{(A) Alanine decamer End-to-end distance probability distribution functions calculated for TRX (black solid line), FISST (blue spheres), REST3 (orange spheres), FISST+REST3 without freezing the weights (red spheres), and FISST+REST3 with freezing the weights (gray spheres) at -10, 0, and 10 pN forces. (B) (left) Corresponding free energy profiles. (right) Free energy scatter plots comparing FISST+TRX and TRX data without and with freezing of the weights.}
\label{fig:ala10figS2}
\end{figure}
\clearpage
\begin{figure}[ht!]
\centering
\includegraphics[width=\textwidth]{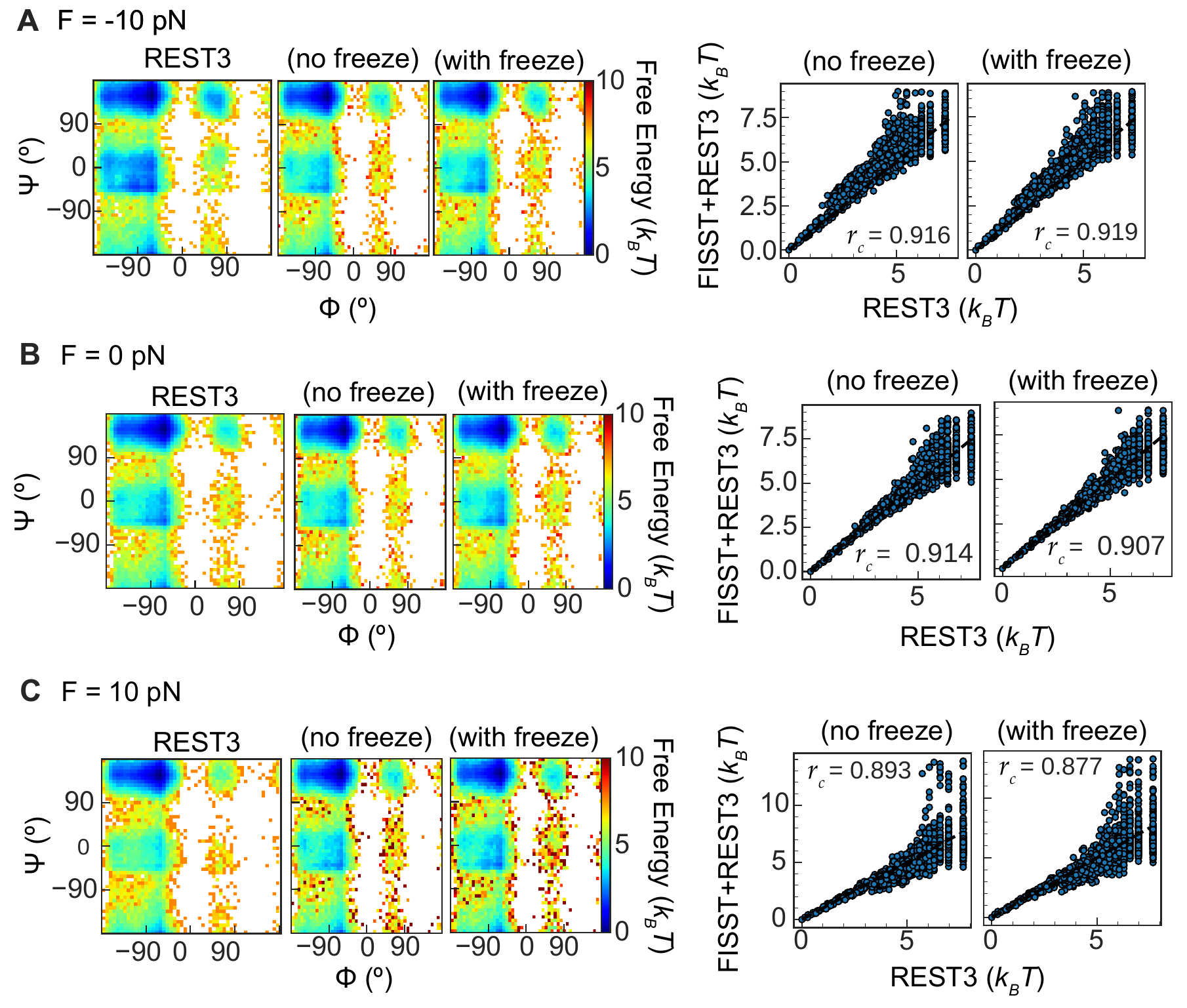}
\caption{(left to right) Ramachandran plots calculated from REST3, FISST+REST3 without freezing the weights, FISST+REST3 with freezing the weights and corresponding free energy scatter plots comparing FISST+REST3 without and with freezing of the weights for (A) -10, (B) 0, and (C) 10 pN forces.}
\label{fig:ala10figS3}
\end{figure}
\clearpage
\subsection{Aib$_9$}
\begin{figure}[ht!]
\includegraphics[width=0.6\columnwidth]{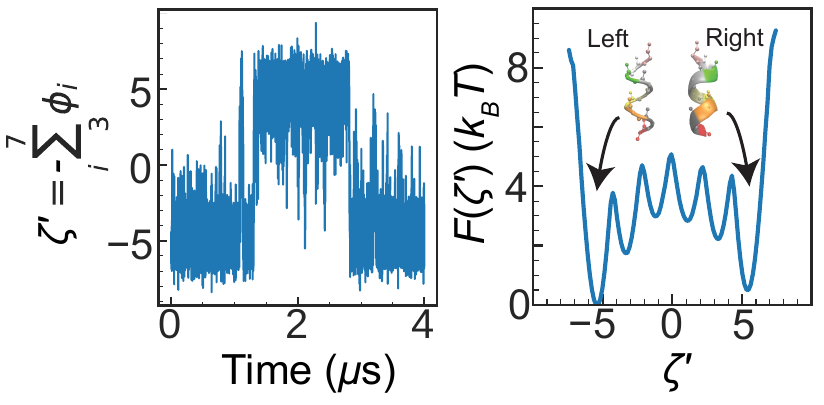}
\caption{(left) Time series plot for $\zeta'$ coordinate generated from unbiased MD simulation of Aib$_9$. (right) Corresponding free energy profile of $\zeta'$. (inset) Left and right-handed Aib$_9$ helices are marked in their respective basins.}
\label{fig:AIB9figS1}
\end{figure}
\begin{figure}[ht!]
\includegraphics[width=0.7\columnwidth]{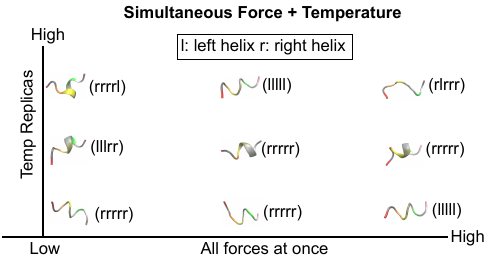}
\caption{Snapshots from Aib$_9$ FISST+REST3 trajectory depicting helical compositions at different forces and solute temperature replicas.}
\label{fig:AIB9figS2}
\end{figure}
\clearpage

\begin{figure}[ht!]
\includegraphics[width=0.4\columnwidth]{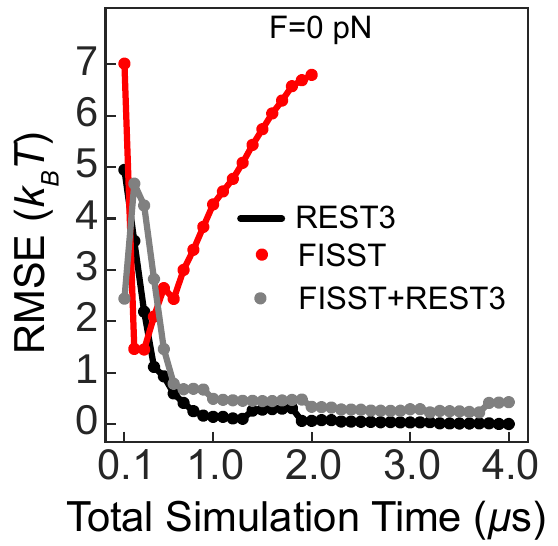}
\caption{RMSE
of $F(\zeta')$ at zero force calculated from full free energy profiles from REST3 (black), FISST (red), and FISST+REST3 (gray) using data
points from different simulation time windows.}
\label{fig:AIB9figS3}
\end{figure}

\begin{table}
     \centering
     \begin{tabular}{ccc}
         & Simulation Time & Wall time \\
       & ($\mu$s) & (days:hours) \\
      \hline
      Unbiased MD & 4.0 & 14:19\\
      \hline \\
      FISST & 2.0&8:9\\
      (single MD)&&\\
      \hline \\
      RX & 0.4 ($\times$10)& 1:16\\
      \hline \\
      FISST+RX & 0.4 ($\times$10)& 1:13\\
      \hline
     \end{tabular}
     \caption{Summary of total simulation and wall times of all Aib$_9$ runs.}
     \label{tab:AIB9comptime}
 \end{table}
 \clearpage
\subsection{Villin (NLE/NLE) Mutant}
\begin{figure}[ht!]
\includegraphics[width=0.8\textwidth]{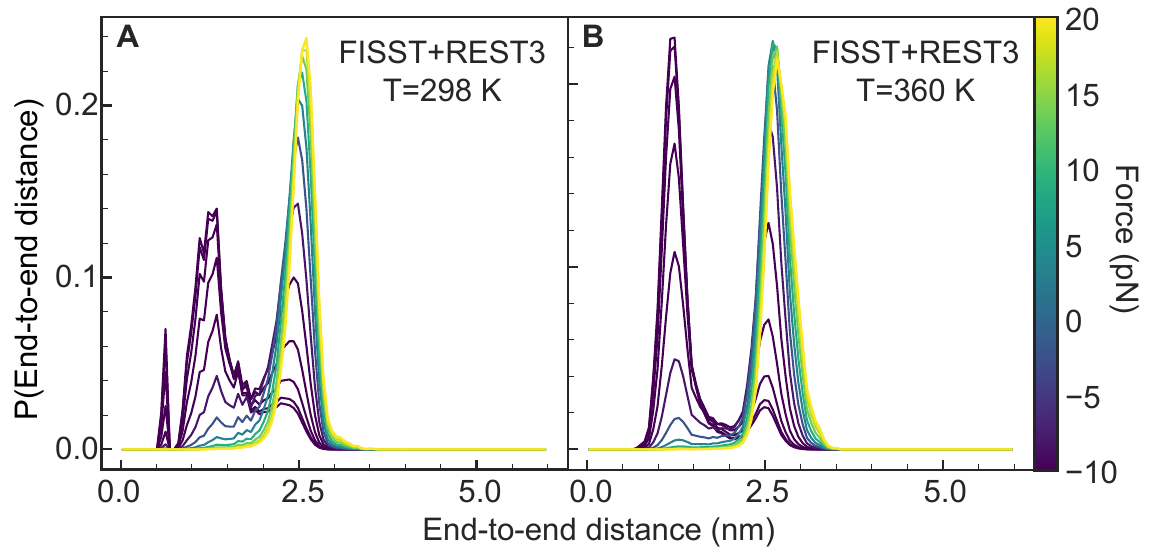}
\caption{Villin Mutant End-to-end distance probability distribution reweighted to forces in the range [-10 pN:20 pN] from FISST+REST3 simulations at (A) Solute temperature of 298K and (B) Solute temperature of 360K. The color bar distinguishes the forces to which the distributions are reweighted.}
\label{fig:hp35figS1}
\end{figure}

\end{document}